\documentclass{aa}
\usepackage{graphicx}
\begin{document}
    
\title{The anomalous Cepheid XZ Ceti\thanks{Based on data obtained at the Mount Stromlo and Siding Spring Observatories, Australia}$^,$\thanks{Tables~1 and 2 are only available in electronic form at http://www.edpsciences.org.}}

\author{L\'aszl\'o Szabados\inst{1} \and 
L\'aszl\'o L. Kiss\inst{2}
\and Aliz Derekas\inst{2}}

\institute{Konkoly Observatory of the Hungarian Academy of Sciences,\\ H-1525 Budapest XII, P.O. Box 67, Hungary
\and School of Physics A28, University of Sydney, NSW 2006, Australia}

\offprints{L. Szabados, \email{szabados@konkoly.hu}}

\date{Received  / Accepted }
                                                                                
\abstract
{}
{
XZ~Ceti is the only known anomalous Cepheid in the Galactic field. 
Being the nearest and brightest such variable star, a detailed study of XZ~Ceti may shed light on the behaviour of anomalous Cepheids whose representatives have been mostly detected in external galaxies.
}
{
CCD photometric and radial velocity observations have been obtained.
The actual period and amplitude of pulsation were determined by Fourier analysis. The long time scale behaviour of the pulsation period was studied by the method of the $O-C$ diagram using the archival Harvard photographic plates and published photometric data.
}
{
XZ~Ceti differs from the ordinary classical Cepheids in several respects. Its most peculiar feature is cycle-to-cycle variability of the light curve. The radial velocity phase curve is not stable either. The pulsation period is subjected to
strong changes on various time scales including a very short one. The ratio of amplitudes determined from the photometric and radial velocity observations indicates that this Cepheid performs an overtone pulsation, in accord with the other known anomalous Cepheid in our Galaxy, BL~Boo (V19 in the globular cluster NGC~5466).
}
{
Continued observations are necessary to study the deviations from regularity, to determine their time scale, as well as to confirm binarity of XZ~Ceti and to study its role in the observed peculiar behaviour.}
{}

\titlerunning{The anomalous Cepheid XZ Ceti} 
\authorrunning{Szabados et al.}
                                                                                
\maketitle
                                                                                
\section{Introduction}
\label{Sect_1}

Variable stars classified as anomalous Cepheids are neglected objects from an observational point of view. Although the first representatives of such variables have been known for more than half a century, the available scanty data are insufficient for the proper explanation of their behaviour. The distinctive features of these variable stars are the short pulsational period (in the period range of RRab stars) and -- strangely enough -- up to 2 magnitudes higher luminosity than that of the RR~Lyrae type stars of the corresponding period. The higher luminosity  relates them to Cepheids, and this is the origin of the term ``anomalous'' Cepheid coined for such variables by Zinn \& Searle (\cite{ZS76}). Its first representatives were found in the Sculptor dwarf galaxy (Thackeray \cite{T50}) and somewhat later in the Ursa Major (van Agt \cite{vA67}) and Draco dwarf galaxies (Baade \& Swope \cite{BSw61}). From the 1970es dozens of anomalous Cepheids have been discovered in various dwarf galaxies of the Local Group. The list of known anomalous Cepheids has been compiled by Nemec et al. (\cite{NNL94}) and more recently by Pritzl et al. (\cite{PAJD02}).

Importance of anomalous Cepheids is twofold. On the one hand, comparison with classical Cepheids may shed light on the differences in the internal structure and evolutionary phase of the two types of pulsators. On the other hand, anomalous Cepheids follow a distinct period-luminosity relationship (see Pritzl et al. \cite{PAJD02} for its most recent form), therefore they serve as standard candles in distance determination for extragalactic systems with no recent star formation.

Existence of pulsating stars in such unusual region in the H-R diagram has been explained in two different ways: (i) by mass transfer (and possibly coalescence) in a close binary system producing a more massive and luminous star than other pulsators on and above the horizontal branch (Zinn \& Searle \cite{ZS76}; Renzini et al. \cite{RMS}) and (ii) assuming that anomalous Cepheids are not old stars but their peculiarity is the extremely low-metallicity (Norris \& Zinn \cite{NZ75}; Demarque \& Hirshfeld \cite{DH75}). The first reliable model for supporting this latter explanation was calculated by Bono et al. (\cite{BCSCP}) with confronting evolutionary tracks and pulsational calculations. This nonlinear convective pulsational model was updated by Marconi et al. (\cite{MFC04}), and these new calculations resulted in a good agreement with the observed behaviour (period, amplitude, colour, luminosity) of the anomalous Cepheids in dwarf spheroidal galaxies.
Two recent studies, an observational (Dolphin et al. \cite{Detal02}) and a theoretical one (Caputo et al. \cite{CCDFM}) explicitly declare that anomalous Cepheids occur in the region which is the extension of the normal Cepheid sequence toward higher temperature for extreme low metallicity ($Z = 0.0002$ to 0.008). Spectroscopic confirmation of this explanation is, however, very difficult, because the extragalactic anomalous Cepheids
are very faint stars for a detailed spectroscopic study.

However, anomalous Cepheids occur in our own Galaxy, in metal-poor
environment. Zinn \& Dahn (\cite{ZD76}) identified the first such variable:
the star V19 in the globular cluster NGC~5466. Curiously, a conventional variable star name, BL~Bootis, has been assigned to this star, though variable stars belonging to globular clusters are catalogued separately. Due to its membership in a globular cluster, luminosity (and luminosity difference with respect to the RR~Lyrae type variables in the same system) of BL~Boo could be reliably determined, and it is this star which advanced to be the prototype of anomalous Cepheids. The BLBOO type variable stars are thus identical with the anomalous Cepheids, and this designation was introduced in the fifth volume of the General Catalog of Variable Stars (Samus \cite{S95}) listing the extragalactic variables. The first detailed spectroscopic analysis ever made on an anomalous Cepheid was that of BL~Boo (McCarthy \& Nemec \cite{McN97}), a 15th magnitude star.

Various pieces of evidence suggest that a much brighter star in the galactic field, XZ~Ceti, also belongs to anomalous Cepheids. In this paper the behaviour of XZ~Ceti is discussed based on new photometric and radial velocity data. Following the overview of the available information on XZ~Ceti (Sect.~\ref{Sect_2}) our observational data
are described (Sect.~\ref{Sect_3}). Then we discuss the value of the pulsation period and its changes, issues related to the pulsation amplitudes of XZ~Ceti and its possible binarity (Sect.~\ref{Sect_4}). Our results are briefly summarized in Sect.~\ref{Sect_5}.

\section{XZ Ceti}
\label{Sect_2}

Photometric variability of XZ~Ceti (HD~12293) was first detected by Hoffmeister (\cite{H33}). A detailed analysis of the Sonneberg photographic plates led Meinunger (\cite{M65}) to conclude that XZ~Ceti was an RR~Lyrae type variable with a period of 0\fd451. This period, however, turned out to be wrong, and Dean et al. (1977) determined the correct periodicity of 0\fd8231 from the first photoelectric photometry of XZ~Ceti.

If XZ~Ceti were really an RR~Lyrae type variable, the 0\fd8231 long pulsation period would imply fundamental mode pulsation and a corresponding RRab type asymmetric light curve. The oscillations of XZ~Ceti, however, result in nearly sinusoidal light variations resembling those of RRc type variables. This peculiarity stimulated a further thorough study carried out by Teays \& Simon (\cite{TS85}). They obtained a photoelectric light curve in Johnson's $B$ and $V$ bands covering a time interval as short as a week. Based on their precise light curve, Teays \& Simon confirmed the value of the pulsation period, and were able to study the shape of the light curve in a quantitative manner, with the help of the Fourier coefficients. In addition, Teays \& Simon (1985) determined the energy distribution from spectrum scans which enabled them to derive the temperature and approximate surface gravity of XZ Ceti.

From a quantitative analysis of the light curve (resemblance of the period, amplitude, and Fourier coefficients to those of BL~Bootis itself), the estimated temperature (6450\,K), surface gravity ($\log g = 2.0$), and their own pulsation models, Teays \& Simon (\cite{TS85}) suggested the possibility that XZ~Ceti is an anomalous Cepheid.

Although XZ~Ceti would be a promising target for a detailed study with its apparent brightness of about 9.6 magnitude in V, it has not been observed purposely since then. Two major sky surveys, however, supplied photometric data on XZ~Ceti: Hipparcos (ESA \cite{ESA97}) and ASAS (Pojmanski \cite{P02}). 

Fortunately, XZ~Ceti was among the targets of RAVE, the most ambitious radial velocity project till now, and its radial velocity value has been made available through the 1st data release (Steinmetz et~al.~\cite{RAVE06}): $204.1\pm 1.2$ km\,s$^{-1}$ at JD\,2\,452\,889.213.

Since XZ~Ceti had been considered as an RR~Lyrae type variable for long (and is still classified as an RRab star in the GCVS!), it was included in various surveys of RR~Lyrae variables. In a major study of kinematics and metallicity of 300 Galactic RR~Lyrae type stars, Layden (1994) published the values 167$\pm$10 km/s for the radial velocity of XZ~Ceti and [Fe/H] = $-2.27 \pm 0.13$ for its metallicity. In another study, Fernley \& Barnes (\cite{FB97}) determined ${\rm v_{\rm rad}} = 190 \pm 10$ km/s and [Fe/H] = $-2.10 \pm 0.13$, in a reasonable agreement with Layden's results for XZ~Ceti. It is to be noted that metallicity XZ~Ceti corresponds to that of the most metal deficient RR~Lyrae type variables. 

\section{New observational data}
\label{Sect_3}

In order to have a deeper insight into the behaviour of XZ~Ceti,
new photometric and spectroscopic observations were performed in 2004-2005.

\subsection{Photometry}
\label{Sect_3.1}
We made time-series $V$-band observations with the 1.0m telescope of the
Australian National University in Siding Spring on seven nights between 2004 
Dec. 25 and 2005 Jan. 11. The detector was one of the eight 2k$\times$4k
chips of the Wide Field Imager, giving $13\farcm0\times26\farcm0$ field of
view (this corresponds to 0\farcs38/pixel image scale). The
exposure time was between 3 s and 10 s, depending on seeing.

\begin{figure*}
\centerline{\includegraphics[width=6.0cm]{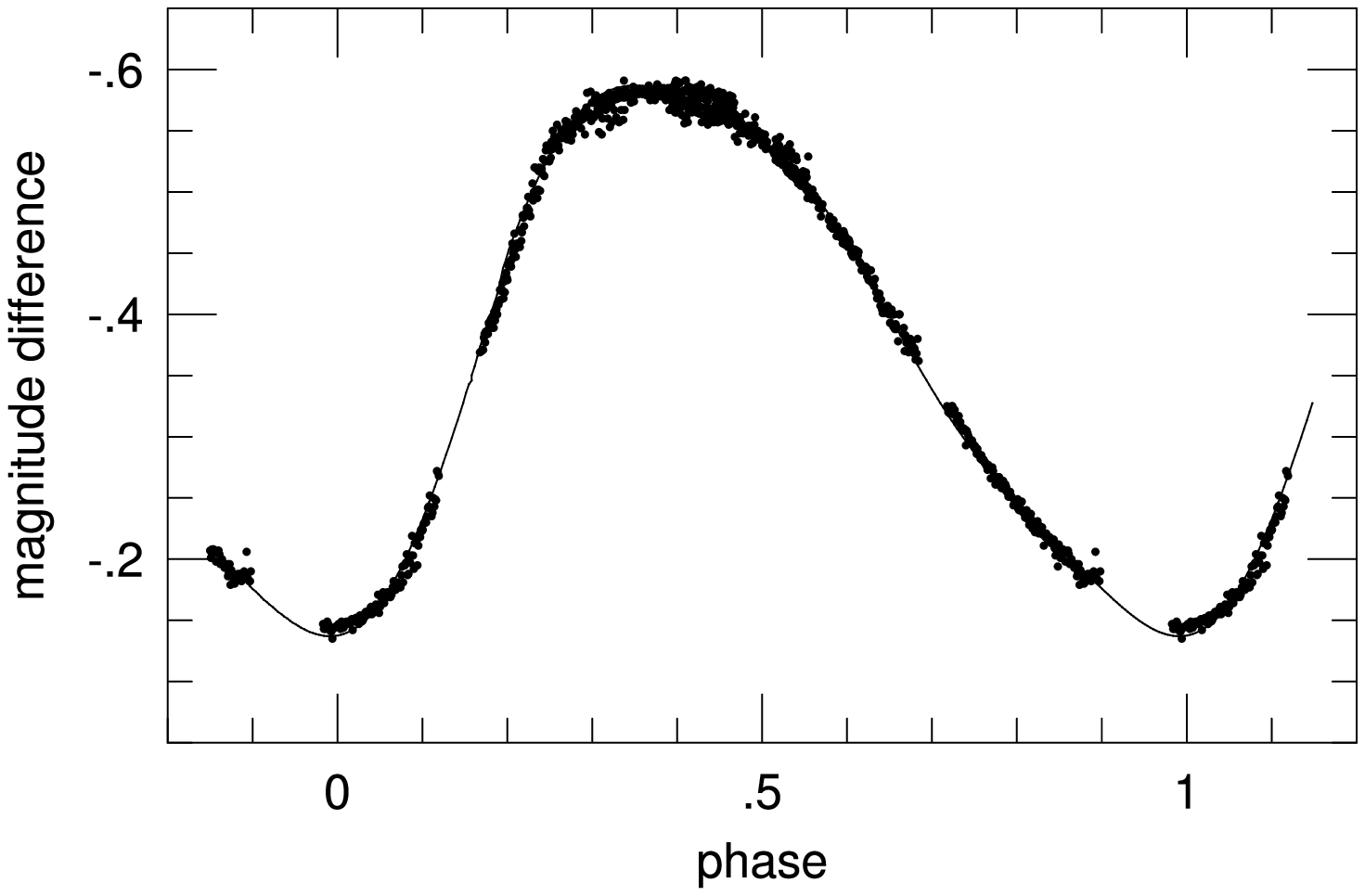}\hspace{2mm}\includegraphics[width=6.0cm]{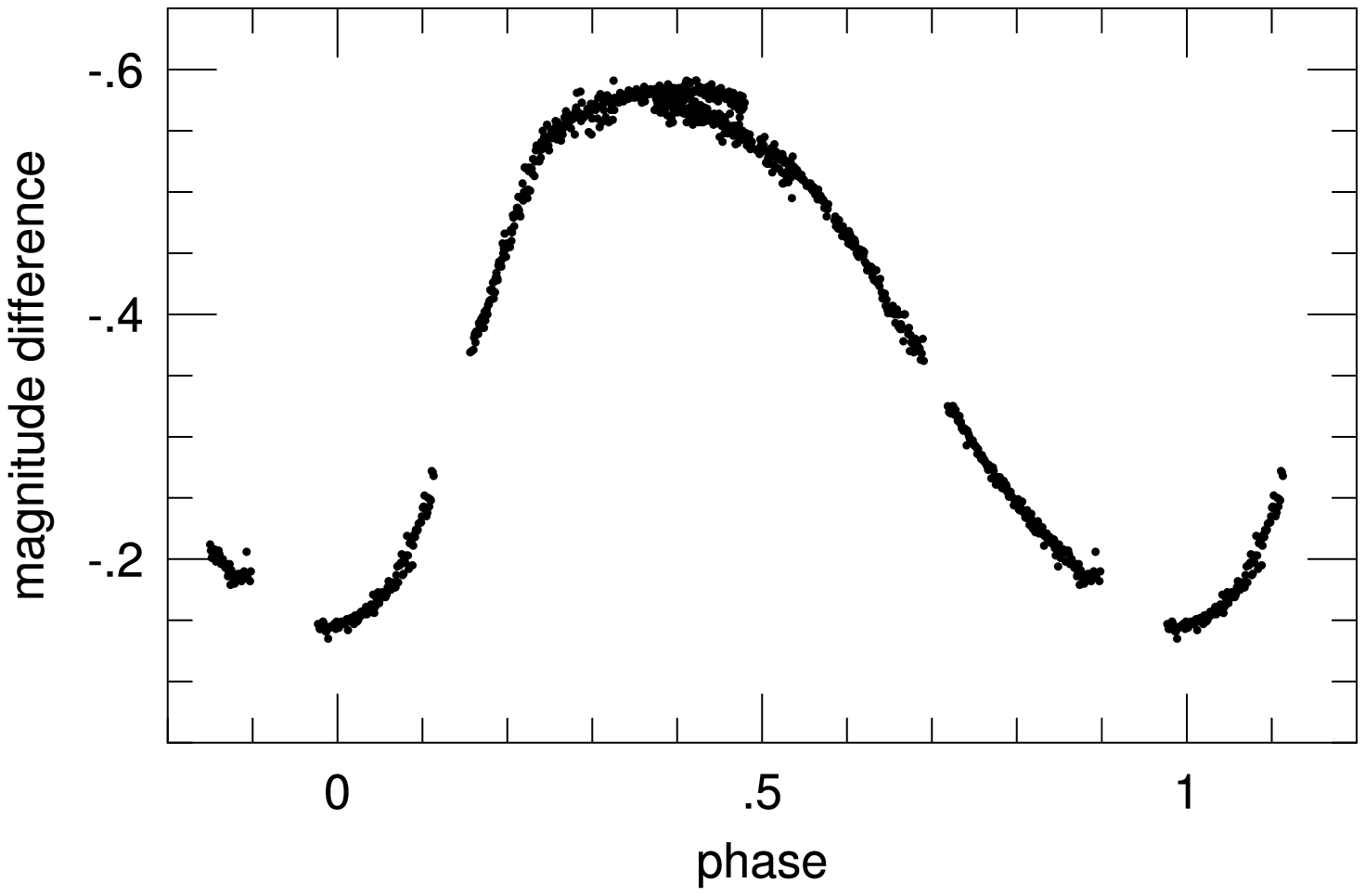}
\hspace{-1mm}\includegraphics[width=6.0cm]{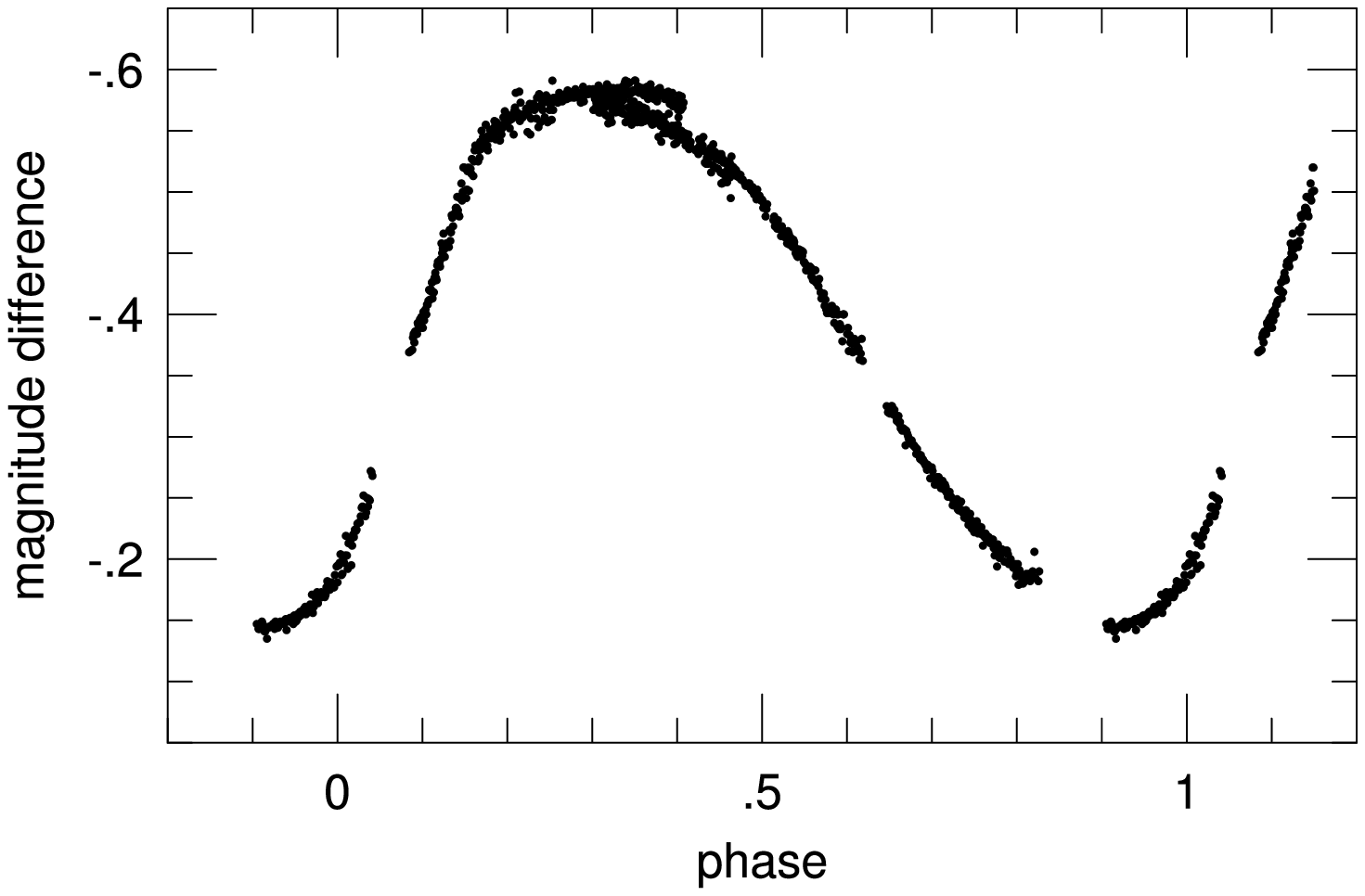}}
\caption{The CCD photometric phase curve of XZ~Ceti in $V$ band. {\sl Left\/}:
The data obtained  have been folded on the best fitting period, 0\fd819. The thin solid line is the Fourier fit involving three harmonics; 
{\sl Middle\/}: The data have been folded on the average period, 0\fd8231298 obtained from the $O-C$ method; 
{\sl Right\/}: The data have been folded on the 0\fd8231561 `instantaneous' period obtained from the straight line fit to the last four $O-C$ residuals. The scatter near maximum brightness still persists.}
\label{Fig_lc}
\end{figure*} 

The observations were reduced with standard tasks in IRAF, using the {\it 
daophot} package. Images were corrected with bias and sky-flat frames, while 
differential magnitudes were determined with simple aperture photometry, 
relative to two local comparison stars, HD~12451 ($V=9.94$) and BD $-16^\circ$350 ($V=10.62$). Typical photometric uncertainty was better than $\pm0.01$ mag, although the relatively large distance between XZ~Cet and HD~12451 (about 23 arcmin) may have 
introduced slightly larger errors on non-photometric nights. The time series of the magnitude differences between the comparison stars testifies that the brightness of both stars is stable.

Because the data were secured in one band (Johnson $V$) only, the magnitudes have not been transformed into the standard system and are treated as differential magnitudes. The existing $BV$ photometry, however, indicates that the difference between the $B-V$ colour indices of the comparison star (HD~12451) and XZ~Ceti is as small as about 0.2 magnitudes, implying that neglect of the colour correction does not have any influence on the results deduced. The photometric observational data (about a thousand points distributed on six nights) are listed in Table~1. The light curve folded on the best fitting period (0\fd819) is seen in the left panel of Fig.~\ref{Fig_lc}. The actual period, however, could be different from this value, as discussed in Section~\ref{Sect_4.1}. (In this paper, zero phase is arbitrarily set at JD~2\,400\,000.0 for all figures.)

\subsection{Spectroscopy}
\label{Sect_3.2}
Spectroscopic observations were carried out with the 2.3m telescope at the 
Siding Spring Observatory, Australia, on four nights between 2004 Dec. 23-29 and during another observational run between 2005 Aug. 17-23. All spectra were taken with the Double Beam Spectrograph using 1200 mm$^{-1}$ gratings in both arms of the spectrograph. 
The projected slit width was 2$^{\prime\prime}$ on the sky, which was
about the median seeing during our observations. The spectra covered the
wavelength range of 5700-6700 \AA, however, we used only 200 \AA\/ centered
on the H$\alpha$ line for radial velocity determination. The exposure time was 90-120 s, 
depending on weather conditions. The dispersion was 0.55 \AA\ px$^{-1}$, 
leading to a nominal resolution of about 1 \AA. In addition to XZ~Cet, we regularly observed telluric and radial velocity standards. 

All spectra were reduced with standard tasks in IRAF. Reduction consisted of 
bias and flat field corrections, aperture extraction, wavelength calibration 
and continuum normalization. We checked the consistency of wavelength 
calibrations via the constant positions of strong telluric features, which 
proved the stability of the system. Radial velocities were determined with 
the task {\it fxcor}, including barycentric corrections. In applying this cross-correlation method, the target spectra were related to the average spectrum of HD~187\,691. This method results in 1-2 km/s uncertainty in the individual radial velocities of XZ~Ceti. Different velocity standards ($\beta$~Virginis, HD~22\,484) have shown that our absolute velocity frame was stable to within $\pm$2--3 km~s$^{-1}$. 

\begin{figure*}
\centerline{\includegraphics[width=6.0cm]{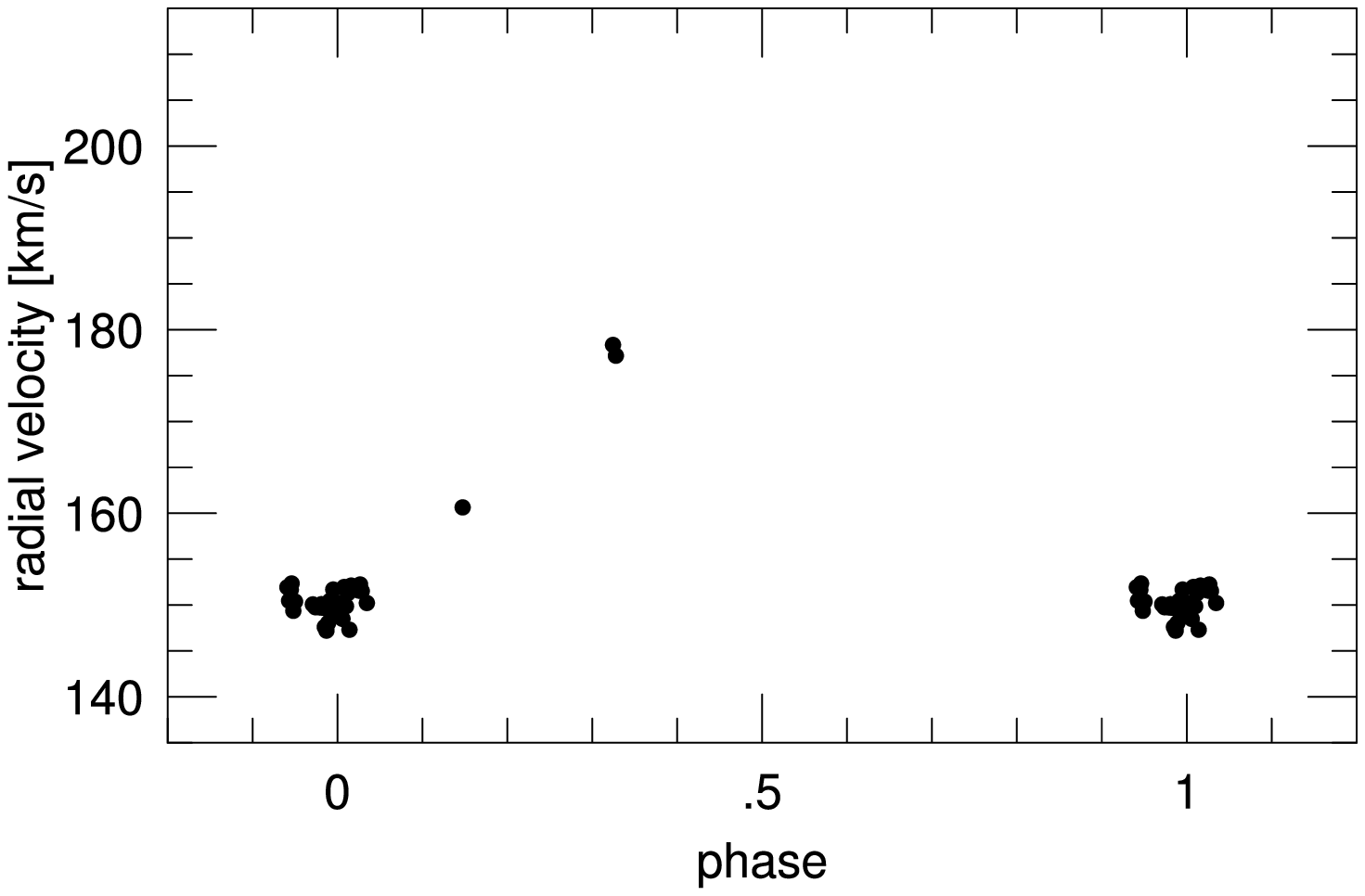}\hspace{0mm}\includegraphics[width=6.0cm]{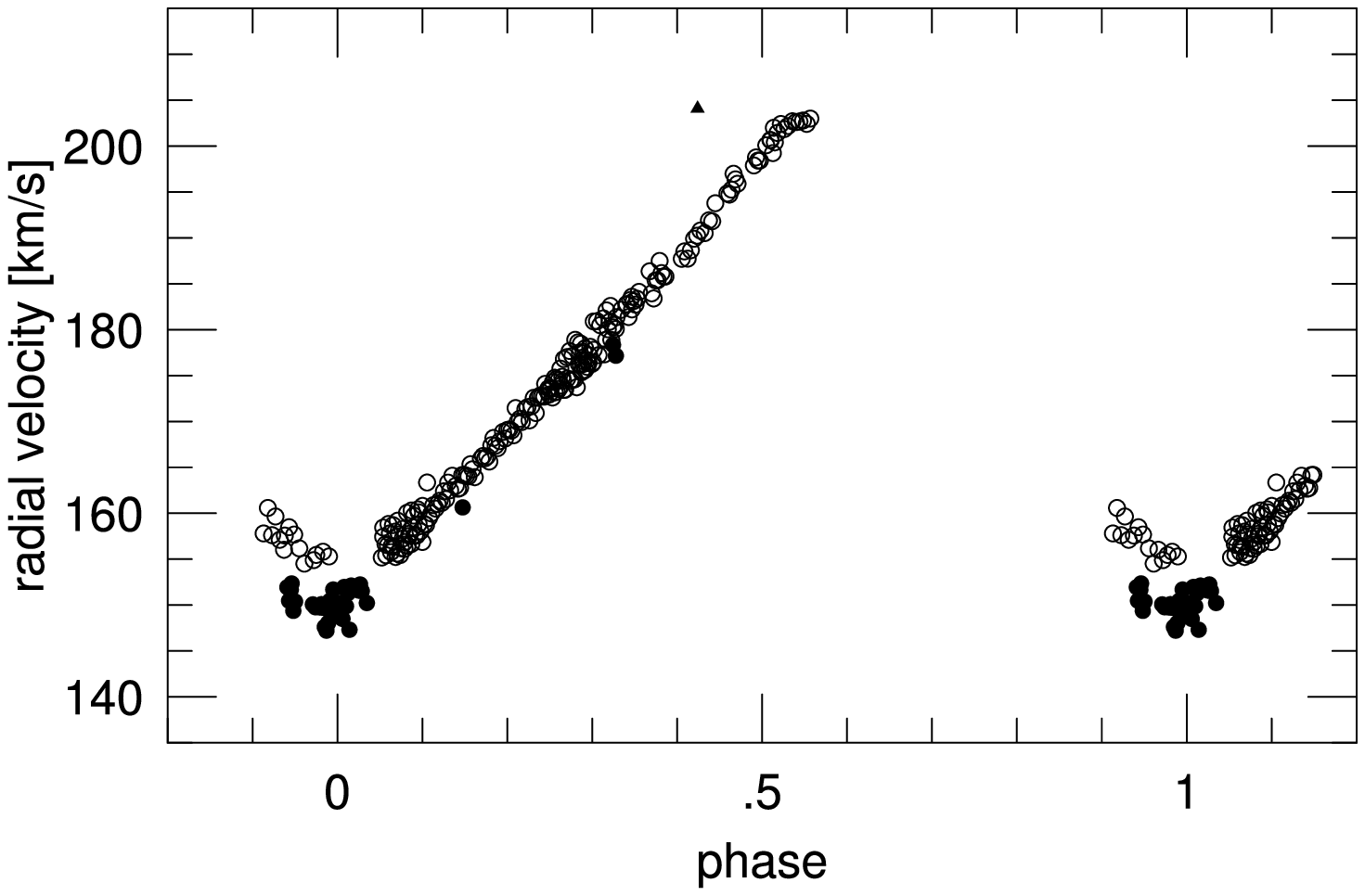}
\hspace{0mm}\includegraphics[width=6.0cm]{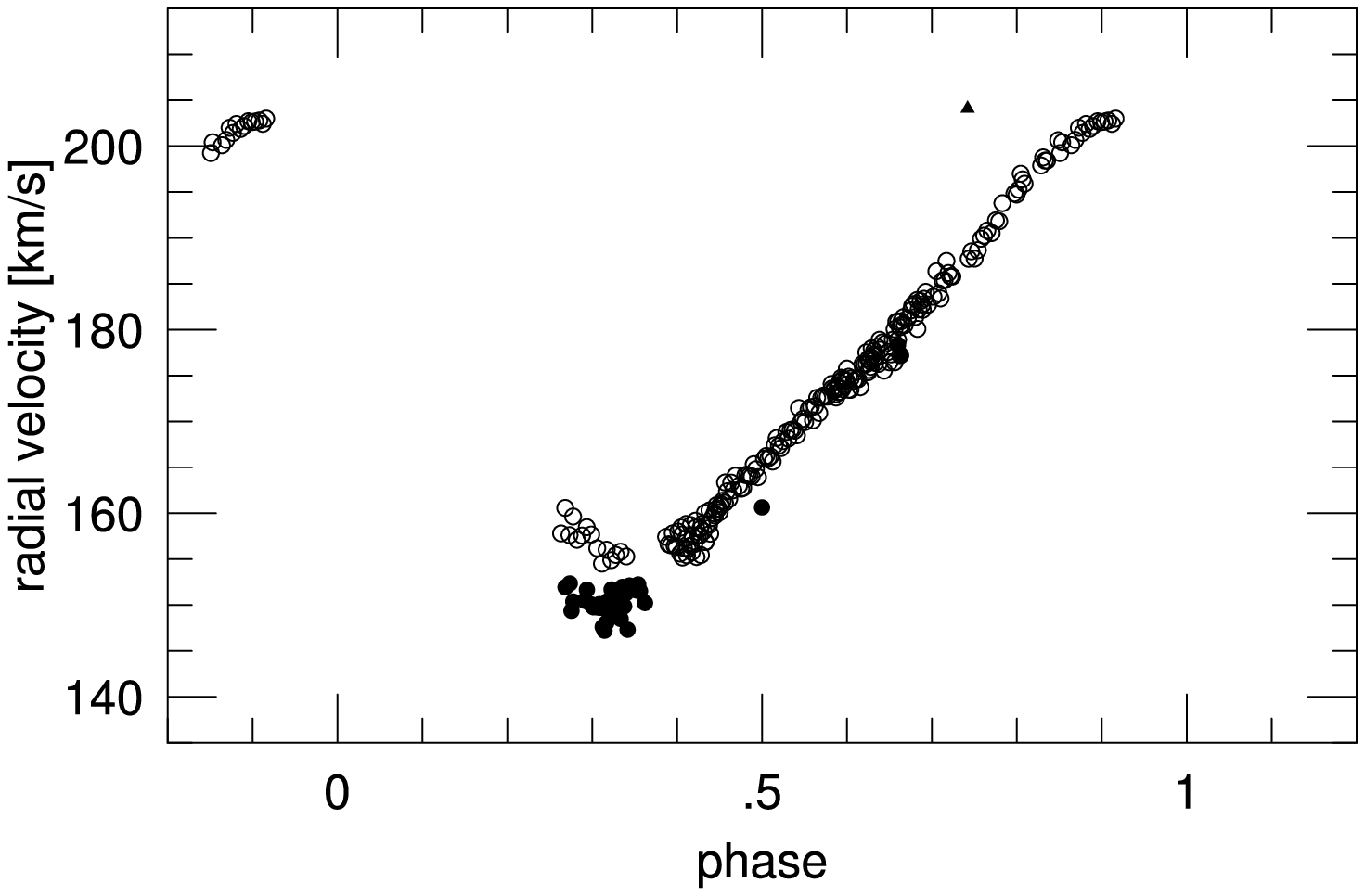}}
\caption{Radial velocity phase curves of XZ~Ceti {\sl Left\/}: The 2004 data folded on the period of 0\fd82604; 
{\sl Middle\/}: The 2004 data (filled circles) and the 2005 data (open circles) plotted together. The data have been folded on the period of 0\fd82604, the value giving the best fit to our radial velocity sample. The single radial velocity value obtained during the RAVE project in 2003 (Steinmetz et~al.~\cite{RAVE06}) is marked by a filled triangle;
{\sl Right\/}: The same as in the middle panel but the data have been folded on the 0\fd8231561 instantaneous period.
The phase curves on the right panels of Figs.~1 and 2 show the actual phasing of the photometric and radial velocity variations.}
\label{Fig_vr}
\end{figure*}

The radial velocity data are listed in Table~2. The data folded on the best fitting period (0\fd82604, determined from the whole radial velocity data set), is seen in the left panel of Fig.~\ref{Fig_vr} for the 2004 data only, while the middle panel of Fig.~\ref{Fig_vr} shows the phase curve constructed from the whole sample. In this latter plot the radial velocities measured in 2004 and 2005 are discerned by different symbols, in order to visualize their deviation from each other. Similarly to the light curve, the actual pulsation period could be different from the value 0\fd82604 obtained from the formal fitting procedure. This suspicion has been confirmed by the available single RAVE data denoted by a filled triangle in Figure~\ref{Fig_vr}.

Though the radial velocity phase curve is not covered completely, the mean radial velocity averaged over the whole pulsational cycle, about 175\,km\,s$^{-1}$, clearly indicates that XZ~Ceti is a Pop.~II object from a kinematical point of view.

\section{Discussion}
\label{Sect_4}

\subsection{Period changes and stability of pulsation}
\label{Sect_4.1}

\subsubsection{The pulsation period}
\label{Sect_4.1.1}

In order to determine the actual value of the pulsation period the program package {\sl MUFRAN\/} was used. This software developed by Koll\'ath~(\cite{K90}) is an efficient tool to point out periodic patterns in various time series. The {\sl MUFRAN\/} software for time-series analysis is a collection of methods for period determination, sine fitting to the observational data, and graphic routines for visualizing the results. Its mathematical basis is the Fourier transform.

The period analysis of our recent dataset gave the smoothest light curve if the actual pulsation period was assumed to be 0.819~d, as is seen in the left panel of Figure~\ref{Fig_lc}. It is noteworthy, however, that the data segments obtained in different nights do not overlap perfectly even if this best fitting period is used which is an evidence of instability of the period and/or light curve. It is to be emphasized that the data have been reduced carefully, and the whole reduction procedure was double-checked when this light curve anomaly was revealed. Adverse effects of bad pixels can be excluded because of the following reason. During the photometric observations we were constantly monitoring the image of XZ Ceti and the two comparison stars in the CCD frames, because being the brightest stars in the field, they could have been close to
saturation even with the shortest exposure times. For that reason, almost
every image was checked with the task {\it imexam} and we can safely exclude the
possibility of being affected by bad pixels -- there were no such pixels in
the vicinity of XZ Cet or the comparison stars. Due to the photometric constancy of the comparison star, the light curve variability of the order of 0.02-0.03 mag is intrinsic to XZ~Ceti.

The instantaneous period can be determined from the radial velocity data,
as well. Although they are relatively less precise than the photometric data, the favourable circumstance that spectroscopic observations were carried out during two runs separated by more than half a year, in principle allows a reliable determination of the pulsation period from the recent radial velocity data. The period search routine of {\sl MUFRAN\/} gave the best fit with a period of 0\fd82604. This value, though considerably differs from that derived from the light curve, seems to be more realistic because of the longer time base. Here, however, another complication emerges: a systematic shift between radial velocities in the subsequent years. An obvious explanation for this feature is that XZ~Ceti belongs to a binary system in which the effect of the orbital motion is superimposed on the pulsational radial velocity changes. The apparently unstable light curve, however, implies that the radial velocity variations are not strictly repetitive, either. This phenomenon and possible binarity of XZ~Ceti is discussed further in Section~\ref{Sect_4.3}.

\subsubsection{$O-C$ diagram}
\label{Sect_4.1.2}

Given this ambiguity concerning the instantaneous pulsation period, the period of pulsation is further investigated using previous photometric data including those obtained by Dean et al. (\cite{DCBW}) and Teays \& Simon (\cite{TS85}). Additional data are available in the databases of Hipparcos (ESA \cite{ESA97}) and ASAS (Pojmanski \cite{P02}). The available precise photometric observations of XZ~Ceti cover about thirty years. Such a long time base is sufficient for studying stability of the pulsation and changes of any origin in the pulsation period. In order to study the behaviour of the pulsation period the usual method of constructing the $O-C$ diagram was applied.

\begin{figure}
\bigskip
\centerline{\includegraphics[width=7cm]{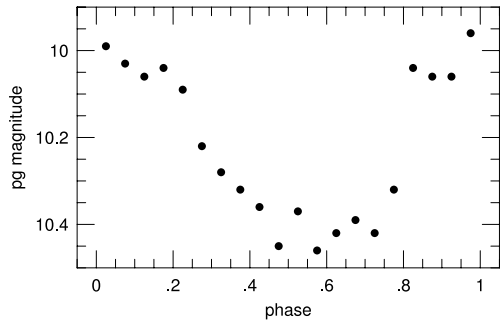}}
\caption{Sample normal phase curve binned from the Harvard photographic data. The curve is based on magnitude estimates covering JD\,2\,427\,000 -- JD\,2\,430\,000 interval.}
\label{pgnor}
\end{figure} 

Furthermore, we could extend the time base by utilizing the Harvard College 
Observatory Photographic Plate Collection. When visiting the 
Harvard-Smithsonian Centre for Astrophysics in early 2006, one of us (LLK) 
obtained photovisual magnitudes of XZ~Ceti on about 2000 photographic plates 
of the archive. XZ~Ceti appears on plates exposed between JD~2\,410\,955 (1888) and JD~2\,447\,771 (1989) and the photographic brightness of XZ Ceti was determined from each plate visually, using local comparison stars. The 
estimated photometric accuracy is about $\pm$0.1 mag per point which, despite the $\sim$0.5 mag full photographic amplitude, allowed us calculating useful normal light curves for shorter segments. After JD~2\,434\,692 the observations of the given celestial region became very sporadic: only 1 per cent of the plates cover the last third of the whole interval. Therefore, we used only photographic data covering the time interval 1888-1953. Omission of the scanty Harvard data from the years 1979-1989 is also justified by existence of more accurate photoelectric data from these years. 

The photographic magnitudes distributed over 65 years were then arbitrarily 
divided into 21 groups, each segment being about three-year-long, and the $O-C$ residual was determined from each normal light curve drawn by about 90 data points on average. A typical binned phase diagram is shown in Fig.~\ref{pgnor}.

In view of the small amplitude of the brightness variation, flatness of the light curve near maximum light, and the possible cycle-to-cycle changes in the light curve, behaviour of the pulsation period of XZ~Ceti has been studied by timing the moments of the median brightness on the ascending branch of the light curve. The phase of occurrence of this feature can be timed more accurately than the phase of maximum brightness, a well proved feature utilized in the $O-C$ analysis of larger amplitude pulsating variables. For smaller amplitude variable stars, like our target, XZ~Ceti, median brightness is reached at the steepest part of the light curve, so its timing has an error as small as about 0.001 in phase, i.e. less than 0\fd001. 

\begin{figure}
\bigskip
\centerline{\includegraphics[width=7cm]{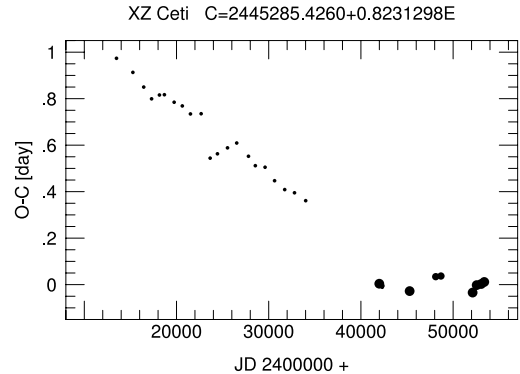}}
\caption{The $O-C$ diagram of XZ Ceti. Size of the circles representing the individual $O-C$ residuals refers to the weight assigned as given in Table~\ref{Tab_oc}.}
\label{Fig_oc}
\end{figure} 

\setcounter{table}{2}
\begin{table}
\caption{$O-C$ residuals of XZ~Ceti with respect to the ephemeris C= JD\,2445285.4260 + 0.8231298$\times$E}
\label{Tab_oc}
\begin{flushleft}
{\small
\begin{tabular}{l@{\hskip1mm}c@{\hskip2mm}c@{\hskip2mm}c@{\hskip2mm}l@{\hskip2mm}c}
\hline
\hline
\noalign{\smallskip}
Norm. median & E  &   O$-$C  &  W & Source \\
J.D.\,2\,400\,000+   &    &  [day]  &  &  \\
\noalign{\smallskip}
\hline
\noalign{\smallskip}
13490.5418 & $-$38628 & +0.9737 &  - &  present paper (pg) \\
15267.6185 & $-$36469 & +0.9131 &  - &  present paper (pg) \\
16443.8077 & $-$35040 & +0.8499 &  - &  present paper (pg) \\
17297.3427 & $-$34003 & +0.7993 &  - &  present paper (pg) \\
18149.2986 & $-$32968 & +0.8158 &  - &  present paper (pg) \\
18687.6267 & $-$32314 & +0.8170 &  - &  present paper (pg) \\
19744.4930 & $-$31030 & +0.7847 &  - &  present paper (pg) \\
20630.1648 & $-$29954 & +0.7688 &  - &  present paper (pg) \\
21510.8792 & $-$28884 & +0.7343 &  - &  present paper (pg) \\
22670.6700 & $-$27475 & +0.7352 &  - &  present paper (pg) \\
23648.3571 & $-$26287 & +0.5441 &  - &  present paper (pg) \\
24440.2266 & $-$25325 & +0.5628 &  - &  present paper (pg) \\
25530.8992 & $-$24000 & +0.5884 &  - &  present paper (pg) \\
26530.1996 & $-$22786 & +0.6092 &  - &  present paper (pg) \\
27805.1707 & $-$21237 & +0.5522 &  - &  present paper (pg) \\
28550.0626 & $-$20332 & +0.5117 &  - &  present paper (pg) \\
29604.4852 & $-$19051 & +0.5050 &  - &  present paper (pg) \\
30636.6320 & $-$17797 & +0.4470 &  - &  present paper (pg) \\
31734.6489 & $-$16463 & +0.4088 &  - &  present paper (pg) \\
32797.2958 & $-$15172 & +0.3951 &  - &  present paper (pg) \\
34011.3782 & $-$13697 & +0.3611 &  - &  present paper (pg) \\
41992.9043 & $-$4000 & $-$0.0025 &  3 & Dean et al. \\
42313.0931 & $-$3611 & $-$0.0112 &  1 & Dean et al. \\
45285.3950 & 0      & $-$0.0310  &  3 & Teays \& Simon \\
48115.3800 & 3438    &  +0.0337  &  2 & Hipparcos \\
48680.0504 & 4124    &  +0.0371 &   2 & Hipparcos \\
52116.5494 & 8299   & $-$0.0308 &   3 & ASAS \\
52556.9564 & 8834    &  +0.0017 &   3 & ASAS \\
53047.5471 & 9430     & +0.0071 &   3 & ASAS \\
53379.2774 & 9833     & +0.0161 &   3 & present paper (CCD)\\
\noalign{\smallskip}
\hline
\end{tabular}}
\end{flushleft}
\end{table}

The individual $O-C$ residuals are listed in Table~\ref{Tab_oc}, and shown plotted in Figure~\ref{Fig_oc} where the size of the circles refers to the weight assigned to the given residual in the fitting procedure. In representing the $O-C$ residuals after JD 2400000, circles of increasing diameters correspond to weights 1, 2, and 3. The Harvard data have been treated separately and with equal weight.

The $O-C$ diagram in Fig.~\ref{Fig_oc} indicates that the period of XZ~Ceti
behaves in a peculiar way: definite short term changes appear in
the period of pulsation. Another obvious feature is that a secular variation also occurs in the pulsation period of XZ~Ceti: in the first half of the 20th century the period was shorter than its average present value. A simple least squares fit to the $O-C$ residuals obtained from Harvard data of XZ~Ceti indicates that the oscillation period was 0\fd8231057 between JD\,2\,413\,000 and JD\,2\,434\,000. The deviations of the $O-C$ residuals from the straight line at about JD\,2\,424\,000 are intrinsic to the pulsation because their values much exceed the uncertainty from the photographic magnitude estimations.

The average value of the pulsation period determined by a weighted least squares fit resulted in the ephemeris:\\
$C$ = JD\,2445285.4260 + 0.8231298$\times$E\\
\phantom{C = JD\,24452\,}$\pm$.0072  $\pm$ 0.0000010

This ephemeris has been valid after JD\,2\,440\,000 but a period jitter is also present in the oscillation of XZ~Ceti throughout the last decades. As a result of this feature, the instantaneous period in a given moment can considerably differ from this value as indicated by the ASAS data and especially by our recent photometry. The best fitting period to the ASAS photometric data is 0\fd82319 by the {\sl MUFRAN\/}, and 0\fd82318 according to the ASAS web-page (http://www.astrouw.edu.pl/$^\sim$gp/asas/asas.html). The practically coinciding values show reliability of the mathematical methods involved.

Although the non-repetitive character of the phase curves (see Sect.~\ref{Sect_4.1.3}) hinders from deriving the precise value of the pulsation period, the $O-C$ diagram itself testifies that period changes have occurred on time scales of years and decades. In this respect the anomalous Cepheid XZ~Ceti differs from classical Cepheids of shortest period because those latter stars pulsate with a more or less stable period. In the case of short period Cepheids only the s-Cepheids have changeable period (Szabados \cite{Sz83}) which stars are thought to pulsate in an overtone mode.

A comparison with the stability of the pulsation period of BL~Bootis would be appropriate. The photometric data of the prototype anomalous Cepheid are, however, of lower quality, so the error bars in the $O-C$ graph constructed by McCarthy \& Nemec (\cite{McN97}) are too large to draw any firm conclusion on the period changes of BL~Boo.

\subsubsection{(In)stability of the pulsation amplitudes}
\label{Sect_4.1.3}

In order to see whether the light curve becomes stable when using a properly chosen pulsation period, the photometric data have been folded on the period of 0\fd8231561, too. This period was obtained by fitting a straight line to the last four $O-C$ residuals, i.e. the ASAS and our photometric data, and it can be considered as the instantaneous value of the period characterising the oscillations in XZ~Ceti. The phase curve plotted with this period (longer than the average value of the period valid for the last decades) is seen in the right panel of Fig.~\ref{Fig_lc}. Again, a definite scatter is seen near the maximum brightness testifying non-repetitive behaviour from cycle-to-cycle.

The radial velocity data have been also folded on the 0\fd8231561 period (see the phase curve in the right panel of Fig.~\ref{Fig_vr}). Although the phase coverage is not perfect, it is clearly seen that the smallest radial velocity values (coinciding with the brightness maxima in phase) are shifted vertically to each other in the subsequent years. If this is a sign of the non-repetitive behaviour of the radial velocity curve, then this effect much exceeds variability of the light curve. Note, however, that our photometry of XZ~Ceti covers only a week, while the time base of our radial velocity data is about half a year. 

The radial velocity value obtained in the RAVE project for XZ~Ceti strongly deviates from its `expected' position in the phase curve. The large shift along the horizontal direction cannot be explained by a considerably different period because this radial velocity observation was obtained at an epoch (JD\,2\,452\,889) when the period behaviour of XZ~Ceti is well known from the ASAS photometry.

Binarity of XZ~Ceti can also explain this shift. In this case orbital motion of the pulsating star around the common mass center causes a long period (corresponding to the orbital period) modulation in the radial velocity variation and the shift occurs in vertical direction. However, intrinsic amplitude changes and the effect of a companion star cannot be separated from the available data. 

\subsection{Pulsation amplitudes and their ratio}
\label{Sect_4.2}
Unlike all other known anomalous Cepheids, XZ~Ceti is not a member in any
stellar aggregation (cluster or external galaxy) whose distance can be determined by some astrophysical method(s), so it is impossible to deduce its position on the colour-magnitude diagram which is necessary to derive its absolute brightness and pulsation mode. The observational data, however, enable us to confirm indirectly that XZ~Ceti is in fact an anomalous Cepheid.

To this end, amplitudes of the photometric as well as the radial velocity variations have been studied which are instrumental in revealing various properties of the pulsating star. The peak-to-peak amplitudes of the brightness and radial velocity 
variations (based on the 0\fd8231561 instantaneous pulsation period) are as follows: 0.450 mag in photometric V-band and 47.28 km\,s$^{-1}$ in the variations of the radial velocity (using the 2005 ${\rm v_{\rm rad}}$ data only). The ratio of the radial velocity amplitude, $A_{\rm v_{\rm rad}}$, and the photometric amplitude, $A_V$, is $A_{\rm v_{\rm rad}}/A_V = 105.1$ km\,s$^{-1}$\,mag$^{-1}$ for XZ~Ceti. It is known that BL~Boo is an overtone pulsator (McCarthy \& Nemec \cite{McN97}, and references therein). From the phase curves shown in the paper by McCarthy \& Nemec, an amplitude ratio of 111.5 can be determined which is even larger than the value for XZ~Ceti.

This amplitude ratio serves as a useful diagnostic tool in classifying pulsating variable stars. For RR~Lyrae type stars this amplitude ratio is 36.4 on average (Liu, 1991). In the case of BL~Her type variables, the photometric and radial velocity data taken from the literature give an average ratio of 47.2. For classical Cepheids, this ratio is 43.6 for fundamental mode and 59.7 for first overtone pulsators (adapted from Szabados \cite{Sz00}). The value exceeding 100 for XZ~Ceti is extremely large to be an RR~Lyrae or BL~Herculis type variable. 

Based on its phenomenological similarity with BL~Boo (shape and amplitude of the light curve and amplitude ratio discussed above), XZ~Ceti is also an overtone pulsator. The amplitude ratio larger than 100 may even indicate second overtone pulsation (Balona \& Stobie \cite{BSt79a}, \cite{BSt79b}).

\subsection{Binarity}
\label{Sect_4.3}
As shown by the radial velocity measurements (Fig.~\ref{Fig_vr}), XZ~Ceti may belong to a spectroscopic binary system. The shift in the radial velocity at a given phase between the 2004 and 2005 data is as large as about $3\sigma$ of the $A_{\rm v_{\rm rad}}$ data, so the systematic difference is significant. A companion decreases the observable photometric amplitude of the pulsating component, while leaves unaffected the corresponding radial velocity variations, thus increasing the amplitude ratio. 

However, one cannot {\it a priori} exclude the pulsation in the second overtone, and in this case an amplitude ratio of about a hundred (twice the value characteristic of fundamental mode pulsation) is expected, corresponding to the frequency ratio of the second overtone and fundamental mode pulsation, $f_2/f_0 \approx 0.5$ for Cepheids. However, the analogy with BL~Boo implies first overtone pulsation for XZ~Ceti. In this case the large amplitude ratio hints at binarity for both XZ~Ceti and BL~Bootis. 

\section{Conclusions}
\label{Sect_5}

We studied photometric and radial velocity variability of XZ~Ceti, a star classified as an anomalous Cepheid. The term ``anomalous'' is obsolete in its original sense, because the recent pulsation models elaborated by Bono et al. (\cite{BCSCP}), Marconi et al. (\cite{MFC04}), and Caputo et al. (\cite{CCDFM}) give a natural explanation why such variables exist in the given region of the H-R diagram: they are extremely metal-poor classical Cepheids. This new paradigm gives motives for comparing behaviour of XZ~Ceti with classical Cepheids.

Resemblance of the phenomenological parameters of the light and radial velocity curves (period, Fourier parameters, amplitude ratios) of XZ~Ceti and the prototype anomalous Cepheid, BL~Bootis corroborates that XZ~Cet is also an anomalous Cepheid. The large value of the mean radial velocity (about 175~km\,s$^{-1}$) is a further piece of evidence that XZ~Ceti does not belong to the same subsystem that includes classical Cepheids.

The systematic shift in the radial velocities at given phases of pulsation hints at possible binarity of XZ~Ceti. Presence of a companion is also inferred from the large radial velocity amplitude as compared with the photometric amplitude, even if greatness of this amplitude ratio is partly caused by the overtone mode of pulsation.

XZ~Ceti remains anomalous in the sense that it shows strong period changes on a very short time scale, and possibly slightly unstable pulsation with cycle-to-cycle variations. Summarizing its longer time scale period behaviour, the characteristic periods have been as follows:\\
between JD\,2\,413\,000 and JD\,2\,434\,000: 0\fd8231057;\\
between JD\,2\,442\,000 and JD\,2\,453\,400: 0\fd8231298;\\ however,
for the interval JD\,2\,452\,000--JD\,2\,453\,400 the most appropriate value is 0\fd8231561.

Further monitoring of this unique variable star is highly desirable.

\begin{acknowledgements}

This project was supported by the Australian Research Council and the Hungarian OTKA grants T042509 and T046207. AD is supported by an Australian Postgraduate Award of the Australian Department of Education, Science and Training. LLK is supported by a University of Sydney Postdoctoral Research Fellowship. The authors are grateful to the anonymous referee for her/his useful comments, Dr. Charles Alcock, the Director of the Harvard College Observatory, for access to the Photographic Plate Collection, Alison Doane, the Curator of the Collection, for helpful assistance, Dr. M\'aria Kun for her critical remarks and careful reading of the manuscript, and Dr. Rachel Moody for retrieving the exact moment of the RAVE observation.

\end{acknowledgements}

\end{document}